\documentclass[twocolumn]{aastex63}
\usepackage{amsmath}

\received{March 22, 2021}
\revised{May 14, 2021}
\accepted{May 17, 2021}

\submitjournal{ApJL}

\shorttitle{COSMOS-dw1}
\shortauthors{Polzin et al.}

\begin{document}

\title{A  recently quenched isolated dwarf galaxy outside of the Local Group environment}

\correspondingauthor{Ava Polzin}
\email{ava.polzin@yale.edu}

\author[0000-0002-5283-933X]{Ava Polzin}
\affiliation{Department of Astronomy, Yale University, New Haven, CT 06511, USA}

\author[0000-0002-8282-9888]{Pieter van Dokkum}
\affiliation{Department of Astronomy, Yale University, New Haven, CT 06511, USA}

\author[0000-0002-1841-2252]{Shany Danieli}
\altaffiliation{NASA Hubble Fellow}
\affiliation{Department of Astronomy, Yale University, New Haven, CT 06511, USA}
\affiliation{Department of Physics, Yale University, New Haven, CT 06520, USA}
\affiliation{Yale Center for Astronomy and Astrophysics, Yale University, New Haven, CT 06511, USA}
\affiliation{Institute for Advanced Study, 1 Einstein Drive, Princeton, NJ 08540, USA}

\author[0000-0003-4970-2874]{Johnny P. Greco} 
\altaffiliation{NSF Astronomy \& Astrophysics Postdoctoral Fellow}
\affiliation{Center for Cosmology and AstroParticle Physics (CCAPP), The Ohio State University, Columbus, OH 43210, USA}

\author[0000-0003-2473-0369]{Aaron J. Romanowsky}
\affiliation{Department of Physics \& Astronomy, One Washington Square, San Jos\'{e} State University, San Jose, CA 95192, USA}
\affiliation{University of California Observatories, 1156 High Street, Santa Cruz, CA 95064, USA}

\begin{abstract}

We report the serendipitous identification of a low mass 
($M_* \sim 2\times 10^6 \, \mathrm{M}_\odot$), isolated, likely quenched dwarf galaxy  in the ``foreground'' of the COSMOS-CANDELS field. From deep 
\textit{Hubble Space Telescope} ({\it HST}) imaging we infer a surface brightness fluctuation distance for 
COSMOS-dw1 of $D_{\mathrm{SBF}} = 
22 \pm 3$ Mpc, which is consistent with its radial velocity of $cz  = 1222 \pm 64$ km s$^{-1}$ via Keck/LRIS. At this distance, the galaxy is 1.4 Mpc in projection from its nearest massive neighbor. We do not detect significant \textsc{H$\alpha$} emission (EW(\textsc{H$\alpha$})$ = -0.4 \pm 0.5$ \AA), suggesting that COSMOS-dw1 is likely quenched. Very little is currently known about isolated quenched galaxies in this mass regime. Such galaxies are thought to be
rare, as there is no obvious mechanism to permanently stop star formation in them; to date there are only four examples of well-studied quenched field dwarfs, only two of which appear to have quenched in isolation. COSMOS-dw1 is the
first example outside of the immediate vicinity of the Local Group.
COSMOS-dw1
has a relatively weak
D$_\mathrm{n}4000$ break and the {\it HST} data show
a clump of blue stars indicating that
star formation ceased only recently.
We speculate that COSMOS-dw1 was quenched due to internal feedback, which was able to temporarily suspend star formation. In this scenario the
expectation is that quenched isolated galaxies with masses $M_*=10^6 - 10^7$\,M$_{\odot}$ generally
have luminosity-weighted ages $\lesssim 1$\,Gyr.

\end{abstract}

\section{Introduction} \label{sec:intro}

Dwarf galaxies in the Local Group play a pivotal role in many areas of astrophysics, including star formation, galaxy formation, and cosmology (see, e.g., \citealt{1998ARA&A..36..435M, 2019ARA&A..57..375S}).
Given their outsized importance and impact there is great interest in finding dwarf galaxies at larger distances,
as they provide information on the environmental dependence of low mass galaxy formation and
can be used to determine how typical Local Group galaxies are for the general population
\citep{2011ApJ...743....8W}.

Most general galaxy catalogs
are biased against the lowest mass dwarfs due to incompleteness  (e.g., \citealt{2008ApJ...688..277T, 2009AJ....137..450W}), which is caused by their 
low surface brightness \citep[see][]{2018ApJ...856...69D}. To overcome this barrier, various telescopes and surveys 
have been designed to be sensitive to low surface brightness emission, making use of specialized algorithms and instruments
to detect/characterize low luminosity galaxies \citep[e.g.,][]{2018ApJ...857..104G, 2020PASP..132g4503V}.
As a result, wide field surveys
are now providing comprehensive catalogs of low luminosity galaxies. 

Past studies of star formation in dwarf galaxies have been subject to these same limitations. For instance, while \citet{2012ApJ...757...85G} were able to study complete samples of galaxies with $M_* \gtrsim 10^8 \; M_\odot$ with SDSS spectroscopy, they probe only a very small volume at lower masses. Galaxies with masses $M_*\lesssim 10^7$\,M$_{\odot}$ have not yet been exhaustively analyzed.

The interpretation of detected low surface brightness
objects generally requires ancillary data, such as
spectroscopy or high resolution imaging (see, e.g.,
\citealt{2021ApJ...908...24G}). 
Archival data offers a shortcut: if dwarf galaxies are sought and found in fields that
already have a suite of ancillary data, the task of determining distances, structural
parameters, and stellar population parameters is far more efficient.
Here we present the serendipitous discovery of an isolated and seemingly quenched dwarf galaxy in the well-studied COSMOS-CANDELS field.

\section{A faint, extended object in the COSMOS field} \label{sec:obj}
\begin{figure*}[ht!]
\epsscale{1.15}
\plotone{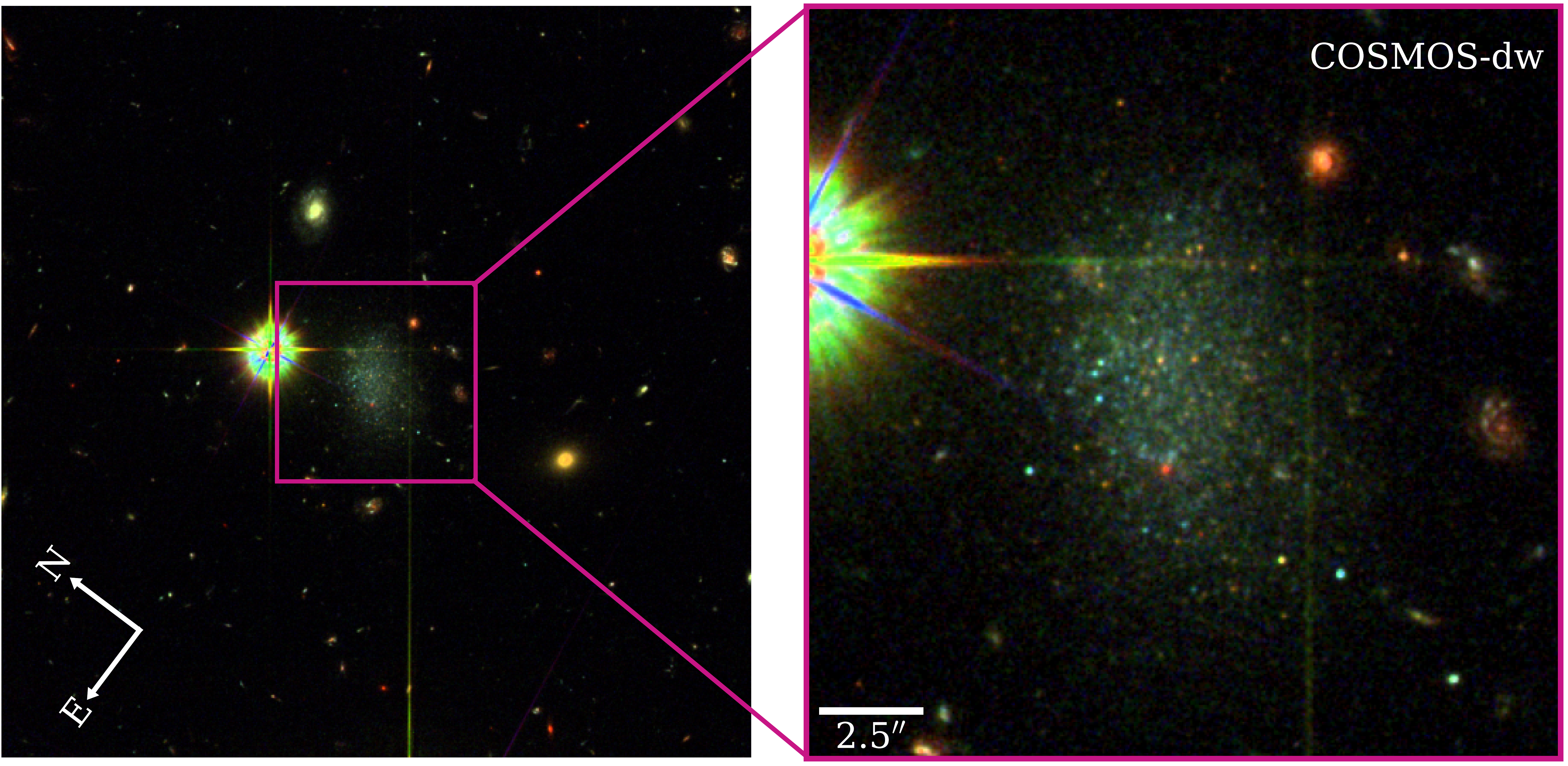}
\caption{Both panels show color composite image of COSMOS-dw1 using F475W, F606W, and F814W. The left panel is $68'' \times 68''$; the right panel is $18'' \times 18''$.
\label{fig:rgb}}
\end{figure*}

\subsection{HST observations} \label{subsec:phot}
The COSMOS-CANDELS \citep{2011ApJS..197...36K, 2011ApJS..197...35G} field covers $\sim200$ square arcmin and is one of the most observed regions in the sky,
with data taken in all major wavelength regimes from X-ray to radio. 

COSMOS-dw1 ($\alpha = 10^h00^m30.03^s$, $\delta = +02^{\circ}08'59.47''$)
was identified in archival \textit{Hubble Space Telescope} (\textit{HST}) data of the COSMOS field  as an object with a semi-resolved appearance, suggesting that it is nearby. The galaxy had been previously overlooked, probably because it is split in multiple very faint objects in standard catalogs \citep{2007ApJS..172...70L, 2015ApJS..219...12A}.

We obtained the CTE-corrected individual {\tt flc} files in the ACS/WFC F475W, F606W, and F814W
bands from the {\it HST} Archive. 
We use \texttt{DrizzlePac}
\citep{2012ascl.soft12011S} to align the images in each filter via \texttt{TweakReg} and then combine them via \texttt{AstroDrizzle}.
\footnote{We did not use existing
data products from CANDELS or 3D-HST for consistency
with our analysis of individual {\tt flc} files in \S\, 4.}
Total exposure times 
are 2028\,s in F475W, 3328\,s in F606W, and 6864\,s in F814W.

An RGB image combining the \textit{HST} data is shown in Figure \ref{fig:rgb}. Several notable features are apparent. COSMOS-dw1 is a low surface brightness, semi-resolved object. Its
appearance is somewhat asymmetric, as it has
a clump of blue stars off-center to the North. The rest of the galaxy
appears to be dominated by red, likely post-main sequence, stars that
are distributed more evenly.

\subsection{\texttt{GALFIT}} \label{subsec:GALFIT}
We begin by measuring the global structural parameters of the galaxy, such as its apparent
size, brightness, and color, using
\texttt{GALFIT} (\citealt{2010AJ....139.2097P}). We first ran \texttt{GALFIT} on the combined F814W+F606W image, after smoothing it with a Gaussian kernel with
$\sigma=2$\,pix ($0\farcs 1$). The galaxy was modeled with a single S\'{e}rsic fit and
the image was aggressively masked, using a \texttt{SEP} (\citealt{1996A&AS..117..393B, Barbary2016}) segmentation map and manually masking additional potential contaminants.
The resulting S\'{e}rsic index is $n=0.69$, and the effective radius is $4\farcs 20$.
Next,  \texttt{GALFIT} was run in each band separately, holding the $r_{\rm eff}$, S\'{e}rsic index,
position angle, ellipticity, and $x,y$ position fixed and letting only the brightness vary in the fit.  
The resultant parameters of our \texttt{GALFIT} runs are listed in Table \ref{table}. Errors are determined in the following way. Eleven copies of the
best-fitting \texttt{GALFIT} model were injected into uncrowded areas in our images and then fitted in the same way as the actual data. The rms variation in the resulting parameters was taken as the uncertainty for each parameter. This method captures errors that are introduced by improper masking of background sources and noise. However, it assumes that the galaxy is smooth and
has a perfect S\'{e}rsic profile, and it does not take deviations from those assumptions into
account.

\begin{deluxetable}{lr}
\tablenum{1}
\tablecaption{COSMOS-dw1 -- Observed properties\label{tab:params}}
\tablewidth{0pt}
\tablehead{parameter & value }
\startdata
$m_\mathrm{{F475W}}$ & $19.43 \pm 0.07$\\
$m_\mathrm{{F606W}}$ & $19.31 \pm 0.04$\\
$m_\mathrm{{F814W}}$ & $19.17 \pm 0.03$\\
F475W$-$F814W & $0.26 \pm 0.08$ \\
F606W$-$F814W & $0.14 \pm 0.05$ \\
$\mu_{0,g}$ (mag arcsec$^{-2}$) & $23.03 \pm 0.08$\\
$\mu_{0,V}$ (mag arcsec$^{-2}$) & $22.90 \pm 0.06$\\
$\mu_{0,I}$ (mag arcsec$^{-2}$)&  $22.77 \pm 0.05$\\
$\mu_{\mathrm{eff},g}$ (mag arcsec$^{-2}$) & $24.19 \pm 0.08$\\
$\mu_{\mathrm{eff},V}$ (mag arcsec$^{-2}$) & $24.07 \pm 0.06$\\
$\mu_{\mathrm{eff},I}$ (mag arcsec$^{-2}$)&  $23.93 \pm 0.05$\\
S\'{e}rsic index & $0.69 \pm 0.01$\\
$b/a$ & $0.721 \pm 0.008$\\
PA (deg) & $-30.2\pm 0.6$\\
$r_{\mathrm{eff}}$ (arcsec) & $4.20 \pm 0.07$\\
$R_{\mathrm{eff}}$ (kpc) & $0.45 \pm 0.06$ \\
$v_\mathrm{rad}$ (km s$^{-1}$)& $1222 \pm 64$ \\
$D_{\mathrm{SBF}}$ (Mpc) & $22 \pm 3$\\
\enddata
\tablecomments{Intrinsic parameters are calculated assuming a distance of 22 Mpc. All magnitudes are quoted in the AB system.} \label{table}
\end{deluxetable}

\section{Velocity and Distance} \label{sec:dist}

 \subsection{Keck Spectroscopy} \label{subsec:LRIS}
\begin{figure*}[ht!]
\epsscale{1.15}
\plotone{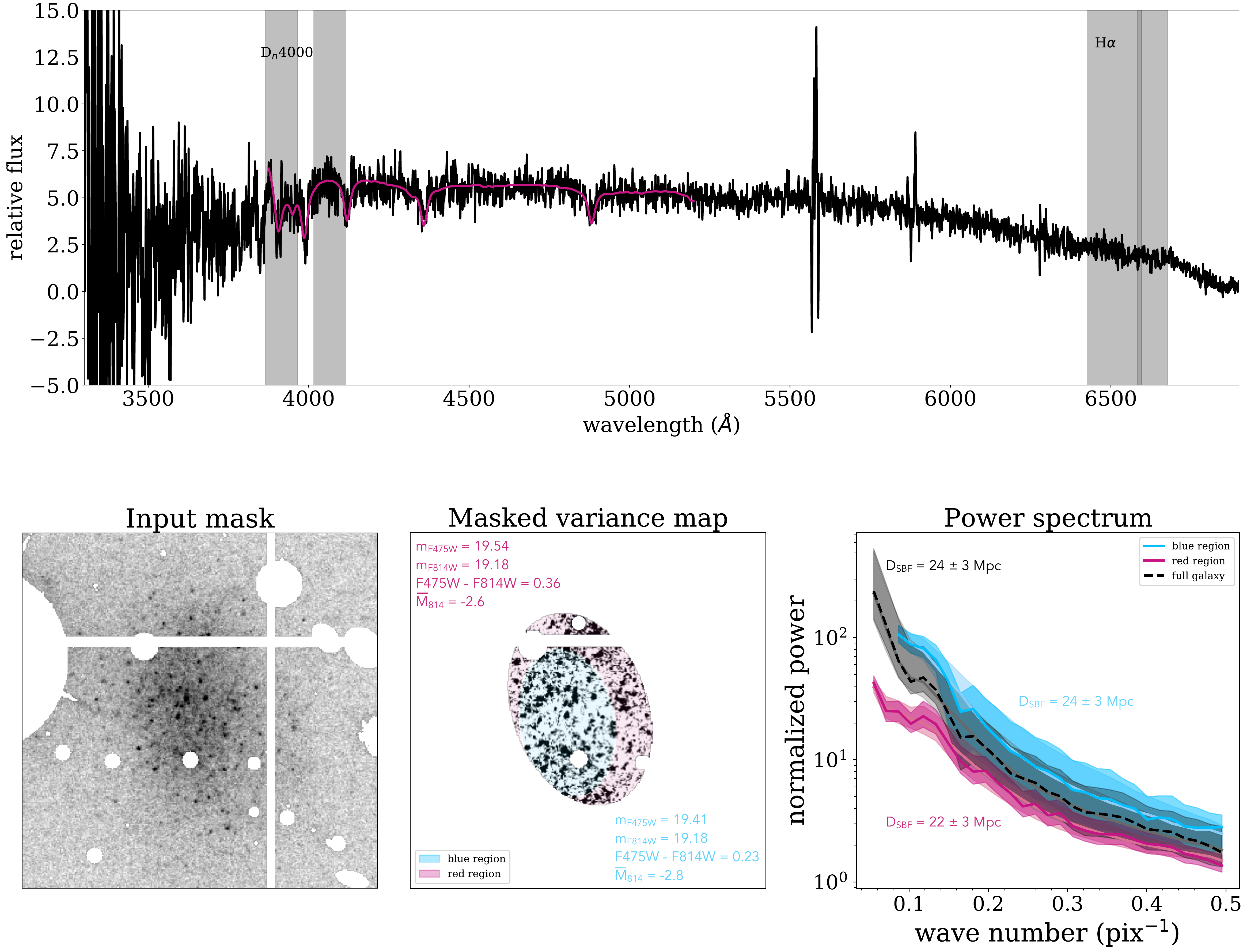}
\caption{Top panel: The median model spectrum (a linear combination of weighted template spectra) $\pm 1\sigma$ output by \texttt{emcee} \citep{2013PASP..125..306F} is shown in pink, plotted over a subset of our LRIS data on the interval 3875 -- 5200 \AA. In black, we show the flux-calibrated spectrum (the instrument response curve was from a different night). Gray regions mark the continuum around the 4000 \AA \, line break -- used to calculate D$_\mathrm{n}4000$ -- and the \textsc{H$\alpha$} line -- used to determine the rms error on the equivalent width. Bottom panel: At left, we show the masked combined F814W image used for the full galaxy SBF analysis (also used when running \texttt{GALFIT}). The middle panel shows the variance map of the full galaxy (as well as the red and blue regions that were analyzed). 
The median of the fits to the power spectra are shown at the right, with the shaded area representing the 68th percentile of the distribution, plotted under the median measured power spectra and the 68th percentile of their distributions. \label{fig:sbf}}
\end{figure*}

We observed COSMOS-dw1 with the Low Resolution Imaging Spectrograph (LRIS, \citealt{1995PASP..107..375O}) on Keck I on 2018 November 5. 
The $1\farcs5$ long slit was used, with the 300\,l\,mm$^{-1}$ grism blazed at 5000\,\AA. 
The total exposure time was 4600\,s in excellent conditions. The data reduction followed standard procedures for long slit data.
The spectrum is shown in Fig.\ \ref{fig:sbf}. The most prominent features are strong Balmer absorption lines, indicating a dominant population of A stars and
an age of $\sim 1$\,Gyr. There are no clearly detected emission lines.

We use a $\chi^2$ minimization scheme over the spectral range $3875 - 5200\,$\AA$\;$ to determine the radial velocity, with \texttt{Flexible Stellar Population Synthesis} (\texttt{FSPS}; \citealt{2009ApJ...699..486C}) template spectra smoothed to the instrumental resolution. We measure a heliocentric radial velocity of $1222 \pm 64$ km s$^{-1}$;
the uncertainty in this result includes fits for a range of ages (0.5--8 Gyr) and metallicities ($-2 \le$ [Fe/H] $\le -1$).

This velocity is consistent with COSMOS-dw1 being at a distance of $20.0\substack{+0.8 \\ -0.7}$ Mpc via the \texttt{Cosmicflows-3} (\citealt{2020AJ....159...67K}; $H_0 = 75 \; \mathrm{km \; s^{-1} \; Mpc^{-1}}$, $\Omega_M = 0.27$, and $\Omega_\Lambda = 0.73$) calculator. However, this uncertainty ignores the peculiar velocity of COSMOS-dw1; a peculiar velocity of 300\,km s$^{-1}$ would correspond to a distance uncertainty of 5\,Mpc.

\subsection{Surface brightness fluctuations} \label{subsec:SBF}

A distance can also be obtained from the \textit{HST} imaging.
The galaxy is only semi-resolved, and we cannot obtain a distance from the
tip of the red giant branch (see Section \ref{sec:DOLPHOT}). 
Instead we use surface brightness fluctuations (SBFs, e.g.,\citealt{1988AJ.....96..807T, 2021ApJ...908...24G}) to constrain the distance to COSMOS-dw1. The SBF method relies on the decreasing pixel-to-pixel brightness variance of a stellar population with increasing distance.

Because this method is sensitive to the nature of the stellar population, we use the integrated galaxy colors shown in Table \ref{table} and the $\bar{M}_{814} \; \mathrm{vs.} \; g_{475} - I_{814}$  relation from \citet{2019ApJ...879...13C}. We find $\bar{M}_{814} = -2.8 \pm 0.3$ using the luminosity-weighted average $g_{475}-I_{814}$ color of the galaxy (see Table 1).

We then generate a variance map from our raw F814W image and the model returned from \texttt{GALFIT} ([image - model]/$\sqrt{\mathrm{model}}$), which we mask, using the aggressive mask we applied to our \texttt{GALFIT} runs plus an elliptical aperture to make sure any measured surface brightness fluctuations are actually coming from the galaxy. In order to avoid biasing our results with selection of aperture size (or the range of assessed wave numbers), we run our SBF analysis repeatedly, randomly selecting our wave number range and ellipse dimensions from reasonable uniform distributions, storing the apparent SBF magnitude from each run. 

As shown in \S\,4.3 the galaxy has a region that is dominated by blue stars. We therefore separately analyze the blue region and red region (in addition to the full galaxy). As is evident from the power spectra (see Figure \ref{fig:sbf}), the full galaxy measurement is strongly affected by the bright, resolved blue stars. As SBF is a more robust method within the redder regime, we adopt the measurement from our red region, finding that the galaxy is located at $22 \pm 3$ Mpc ($\bar{M}_{814} = -2.6 \pm 0.3$ and $\bar{m}_{814} = 29.15 \pm 0.08$), consistent with the redshift distance, which suggests this galaxy has a small peculiar velocity. The results from the full galaxy/blue region fall within these error bars ($24 \pm 3$ Mpc), and the measurements for each subset of the galaxy are also consistent with the SBF distance using an extrapolation
of the \citet{2010ApJ...724..657B} relation.

We determine distance-dependent quantities with $D_\mathrm{SBF}$ (see Table \ref{table}).
We find that the galaxy has a luminosity of $L_{\rm F606W} = (7 \pm 2) \times 10^6 $\,L$_{\odot}$ and
a physical size of $R_\mathrm{eff}= 450 \pm 60$\,pc.

\section{Stellar population}

\subsection{Constraints from the integrated colors} \label{colors}

We first use the \textit{HST}-measured colors to constrain the stellar population properties. To break the age--metallicity degeneracy we assume that COSMOS-dw1 falls on the mass--metallicity relation. Running a grid of ages and metallicities through \texttt{FSPS}, COSMOS-dw1's integrated F606W$-$F814W and F475W$-$F814W colors and absolute F606W magnitude imply a stellar population of age 0.9 Gyr, [Fe/H] = $-1.6$, and $M_* = 2.4\times10^6 \; M_\odot$. 
This age is qualitatively consistent with the prominent Balmer lines in the spectrum.

\begin{figure*}[ht!]
\epsscale{1.15}
\plotone{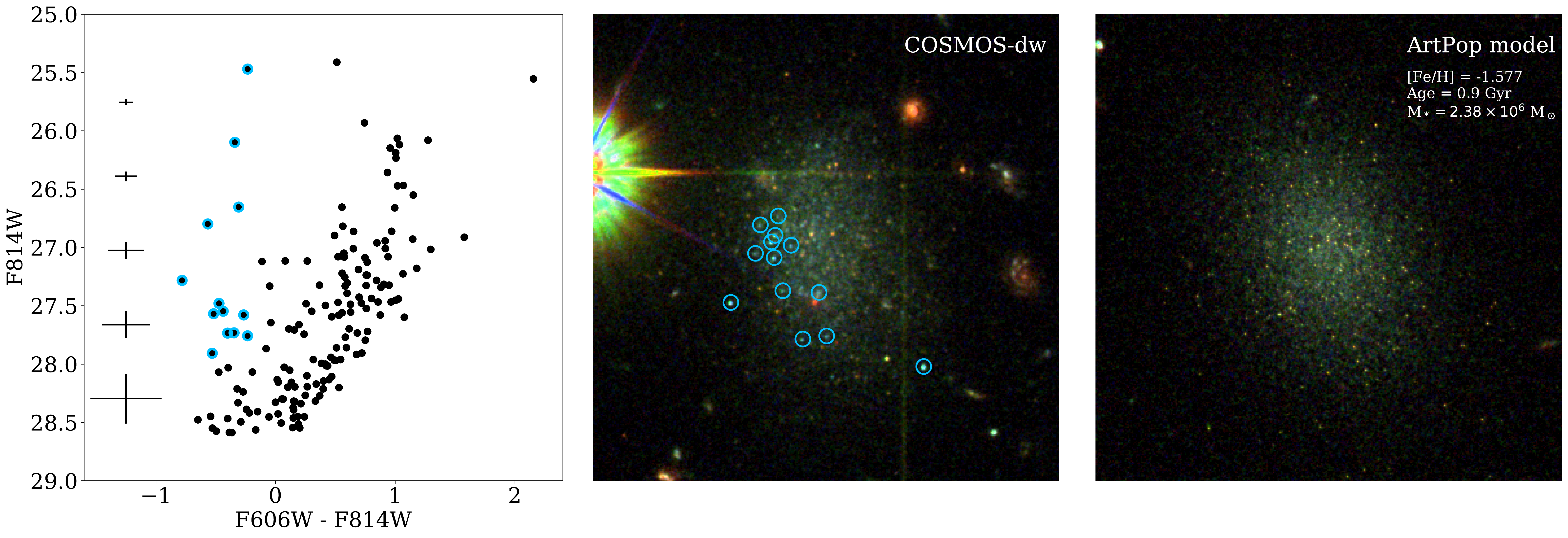}
\caption{Left panel: COSMOS-dw1's CMD is shown using stellar photometry from \texttt{DOLPHOT}. Average error bars for each  m$_{\mathrm{F814W}}$ bin are shown at the far left. The points marked by a blue circle correspond to the brightest, bluest stars with F606W - F814W $\le -0.2$ and m$_\mathrm{F814W} \le -28$. These same stars are marked in RGB image of COSMOS-dw1 in the middle panel. Right panel: An \texttt{ArtPop} model of a simple stellar population placed at 22 Mpc that has the same integrated properties as COSMOS-dw1 (see Table \ref{tab:params}).  \label{fig:CMD}}
\end{figure*}

To illustrate that this SSP provides a reasonable description of the galaxy we use the \texttt{ArtPop} code (first described in \citealt{2018ApJ...856...69D}; J. Greco and S. Danieli, in prep.), which creates full 2D models of galaxies by drawing stars from isochrones. 
We then  inject the simulated galaxy into our drizzled \textit{HST} images on a filter-by-filter basis (see the rightmost panel of Figure \ref{fig:CMD}). The overall appearance is a good match, although the morphological structure of the model is clearly more regular than that of the data. Furthermore, there are blue stars that are not accounted for in the \texttt{ArtPop} model; we will return to those below.

\subsection{Constraints from spectral indices}

We measure EW(\textsc{H$\alpha$}) and the strength of the 4000 \AA \, line break \citep{1999ApJ...527...54B}. We find an equivalent width consistent with no \textsc{H$\alpha$} emission ($-0.4 \pm 0.5$ \AA) and D$_\mathrm{n}$4000 index = $1.22 \pm 0.02$, which is \textit{just} inconsistent with a quenched galaxy per the relation and criterion from \citet{2012ApJ...757...85G}.  
The \textsc{H$\alpha$} equivalent width corresponds to a $3\sigma$ specific star formation rate (sSFR) upper limit of $1.5\times10^{-12} \,  \mathrm{yr}^{-1}$ \citep{2018MNRAS.477.3014B}. We infer that the galaxy is young but is not forming stars at present.

\subsection{Evidence for a complex stellar population} \label{sec:DOLPHOT}

We use the ACS module from \texttt{DOLPHOT}, an adapted version of \texttt{HSTphot} \citep{2000PASP..112.1383D}, to obtain photometry of individual stars in COSMOS-dw1. We follow the module's pre-processing steps, including bad pixel rejection, sky estimation, and fine alignment of the input images. Using our drizzled F814W image as the reference, we run \texttt{DOLPHOT} twice, once for all F814W and F606W \texttt{flc} files and once for F814W and F475W. For the photometry's PSF fitting, we use \texttt{TinyTim} PSFs \citep{2011SPIE.8127E..0JK}.

Our parameter files closely follow the recommendations from the \texttt{DOLPHOT} handbook, but we adopt the \citet{2009ApJS..183...67D} values for the sky-fitting parameter (\textit{FitSky} = 3), aperture radius (\textit{RAper} = 10 pix), and flag that forces all detected sources to be treated like stars for the purpose of fitting (\textit{Force1} = 1).

We make quality cuts on the detected/photometered sources in the \texttt{DOLPHOT} output to make sure we only include sources with high quality stellar photometry. Following \citet{2017ApJ...837..136D}, we include ``good stars'' (\textit{object type} = 1) with high quality photometry (\textit{photometry quality flag} $\le$ 2), high signal-to-noise (\textit{signal-to-noise} $\ge$ 4), and object sharpness within a star-like range ($-0.3 \le$ \textit{sharpness$_\mathrm{F606W+F814W}$} $\le 0.75$). Spatially, we include all stars
that reasonably belong to COSMOS-dw1. The resultant color--magnitude diagram (CMD) is shown in the leftmost panel of Figure \ref{fig:CMD}.

As is notable in the color composite images of COSMOS-dw1, the CMD shows a population of bright, very blue stars. These stars (F606W$-$F814W $\le -0.2$ and $m_\mathrm{F814W} \le -28$) are marked in both the CMD and the RGB image in Figure \ref{fig:CMD}. Their location 
provides an upper limit to their age: the dynamical time at their distance
from the center is only $\approx 10^8$\,yr, and the stars would have
dispersed throughout the galaxy if they formed earlier than that.

\section{Environment} \label{sec:environ}
Dwarf galaxies are thought to be quenched predominantly by environmental effects, such as  ram pressure stripping and tidal stripping (e.g., \citealt{2006PASP..118..517B, 2011ApJ...739....5W, 2018MNRAS.477.4491F}), so it is expected that seemingly quiescent galaxies with little-to-no evidence of \textsc{H$\alpha$} emission are within one to two virial radii of a bright companion. 
Intriguingly, COSMOS-dw1 does not have an obvious luminous companion.

We search within 5$^{\circ}$ ($\sim 2$ Mpc projected at 22 Mpc) and 300 km s$^{-1}$ of COSMOS-dw1 in order to assess its immediate environment. There are 19 nearby galaxies within this projected distance in the radial velocity range 922--1522\,km\,s$^{-1}$. Assuming each of these nearby galaxies is also located at a distance of 22 Mpc and using an \textit{r}-band mass-to-light ratio of 
$\Upsilon_r = 3.05\; M_\odot\, L_\odot^{-1}$ \citep{2003ApJS..149..289B}, we find that two of the 19 galaxies exceed the minimum mass to potentially be considered a ``luminous neighbor'' ($M_\star > 2.5 \times 10^{10}\; M_\odot$) as in \citet{2012ApJ...757...85G}. The closest of these,
NGC\,3166, is 1.4\,Mpc away in projection, just inside the 1.5\,Mpc limit used by Geha et al \footnote{We note that the intriguing early-type galaxy Ark 227 \citep[][]{1975CoBAO..47....3A} is at a projected distance of only 4 arcmin from COSMOS-dw1. However, its redshift ($cz = 1793 \, \mathrm{km \; s^{-1}}$; \citealt{1999PASP..111..438F}) is $571 \, \mathrm{km\; s^{-1}}$ removed from that of COSMOS-dw1, corresponding to a distance of approximately $30\, \mathrm{Mpc}$ and effectively ruling out an association.}.

\begin{figure*}[ht!]
\epsscale{1.15}
\plotone{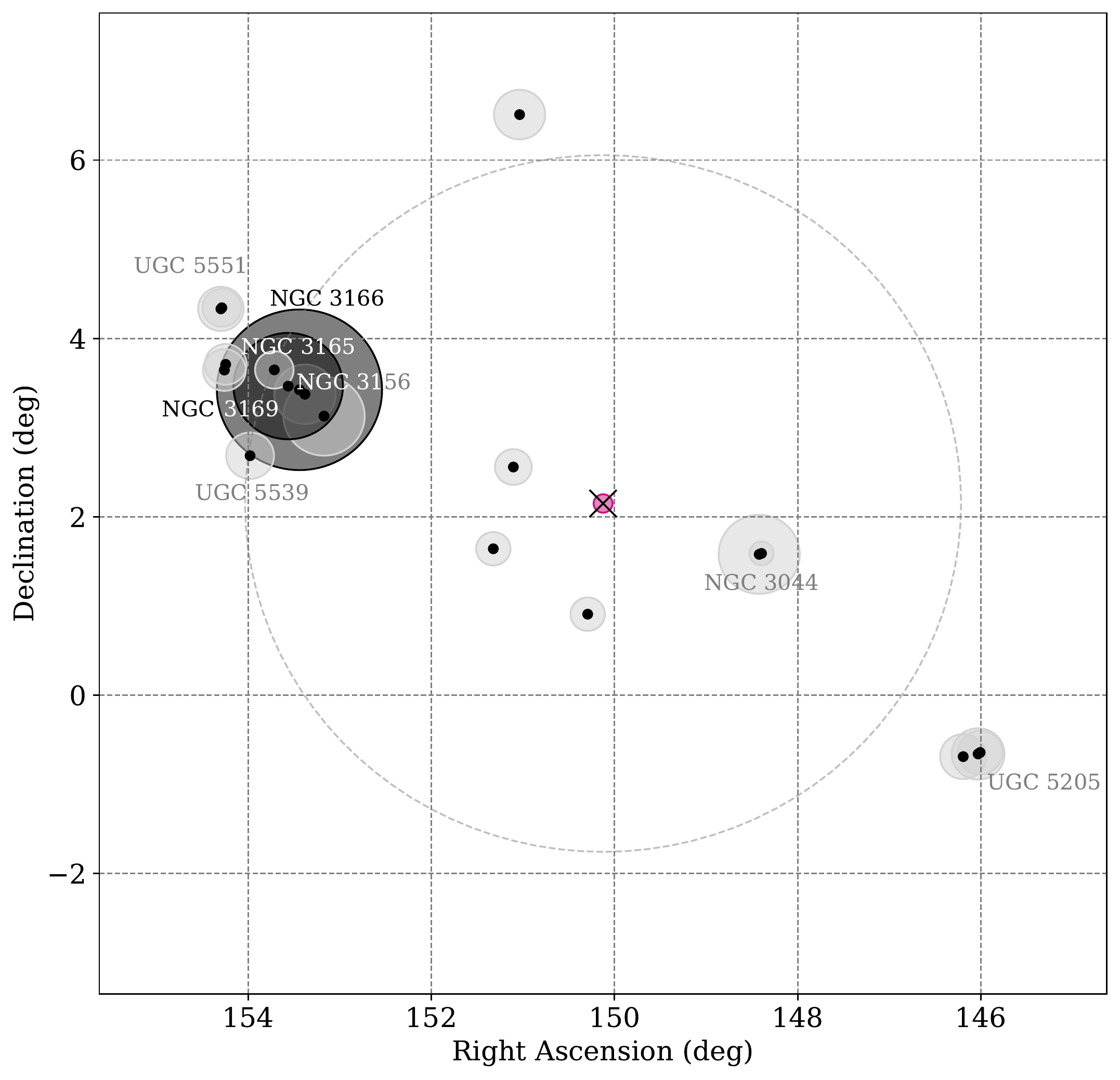}
\caption{We assess the environment of COSMOS-dw1 by estimating the virial radii of nearby galaxies, assuming that they are all at the same distance as the dwarf. We show the 1.5 Mpc projected radius as the dashed gray circle, while all galaxies within 5$^{\circ}$ and 300 km s$^{-1}$ of COSMOS-dw1 are shown as black points with their virial radii underplotted; galaxies with $M_* > 2.5 \times 10^{10} \, M_\odot$ are marked by darker underplotted virial radii.
NGC and UGC objects are labeled. Each square of the grid is $0.77 \times 0.77 \; \mathrm{Mpc}^2$ at a projected distance of 22 Mpc. We find that COSMOS-dw1 is at least $\gtrsim 4$ virial radii away from any other galaxy.\label{fig:environ}}
\end{figure*}

We further explore an association with this galaxy, or other nearby galaxies, by estimating
the virial radii of potential neighbors using
the stellar mass--r$_{80}$ and r$_{80}$--virial radius relations defined in \citet{2019ApJ...872L..13M}, which take $\Delta_c = 200$.
The results are shown in Figure \ref{fig:environ}. 
NGC 3044, the closest non-dwarf galaxy to COSMOS-dw1 with a calculated stellar mass below our luminous neighbor threshold ($M_* \sim 1.2 \times 10^{10} \, \mathrm{M}_\odot$) and $v_\mathrm{rad} - v_\mathrm{rad,dw} = 66 \;\mathrm{km \; s^{-1}}$, is located $4.1R_{\rm vir}$ from our isolated dwarf in projection. NGC 3166, the closest luminous neighbor ($M_* \sim 6.1 \times 10^{10} \, \mathrm{M}_\odot$, $\Delta v_\mathrm{rad} = 106 \;\mathrm{km \; s^{-1}}$), is
$3.9R_{\rm vir}$ away. It is worth noting that these values are strict lower limits, as we assume that the projected distances equal the physical distances.

There are also three dwarf galaxies near COSMOS-dw1: LEDA 1230703 ($M_* = 1.2 \times 10^8 \; M_\odot$, $\Delta v_\mathrm{rad} = -109 \;\mathrm{km \; s^{-1}}$), 0.41 Mpc ($5.3R_{\rm vir}$) from COSMOS-dw1; 2dFGRS TGN353Z197 ($M_* = 8.3 \times 10^7 \; M_\odot$, $\Delta v_\mathrm{rad} = 124 \;\mathrm{km \; s^{-1}}$) at a distance of 0.48 Mpc ($6.7R_{\rm vir}$); and SDSS J100517.67+013831.2 ($M_* = 8.3 \times 10^7 \; M_\odot$, $\Delta v_\mathrm{rad} = 43 \;\mathrm{km \; s^{-1}}$), 0.50 Mpc ($6.9R_{\rm vir}$) from COSMOS-dw1.

We note that the nearest galaxies' $\Delta v_{rad}$ distribution is consistent with the distribution of the full sample returned in our search.

\section{Discussion and Conclusion} \label{sec:disc}

We report the serendipitous discovery of an isolated quenched low mass galaxy at a distance of $\sim 22$ Mpc. All other known and well-studied galaxies in this region of parameter space are in the immediate vicinity of the Local Group. The four isolated Local Group dwarfs are Cetus \citep[][]{1999AJ....118.2767W}, Tucana \citep[][]{1990IAUC.5139....2L}, KKR\,25
\citep{2012MNRAS.425..709M}, and KKs\,3 \citep{2015MNRAS.447L..85K}. There is evidence to suggest that Cetus and Tucana are backsplash galaxies \citep{2012MNRAS.426.1808T} which were environmentally quenched during a previous passage through the Local Group.  KKR\,25 and KKs\,3 are $\approx 2$\,Mpc away from
the Local Group, further than the projected distance of COSMOS-dw1 to its nearest
potential neighbor. However, NGC\,3166 and NGC\,3044 have a lower mass than the Milky Way
and M31, and when expressed in virial radii, KKR\,25 and KKS\,3 are a factor of $\approx 2$ {\em closer} to their nearest neighbor than COSMOS-dw1 is.

The quenching mechanism for COSMOS-dw1 is a puzzle.
The isolation of the galaxy combined with the fact that quenching happened recently makes
an environmental cause very unlikely. Interestingly, all three galaxies (COSMOS-dw1, KKR\,25, and
KKS\,3) have complex stellar populations \citep[][]{2012MNRAS.425..709M, 2015MNRAS.447L..85K}, suggesting that star formation
stopped and started several times over their lifetimes.

We suggest that quenching was due to internal feedback. Simulations suggest that supernova feedback can shut down star formation in
low mass galaxies, but only for a short time \citep{2017MNRAS.471.3547F, 2018MNRAS.477.4491F}. Whenever we catch a galaxy in this short-lived phase we observe
it to be young. Of the three galaxies in this limited sample, COSMOS-dw1 appears to be the youngest, and the clump of blue stars in COSMOS-dw1 may represents the site of the
feedback event that temporarily halted further star formation \citep[see also][]{2012ApJ...757...85G}. The intriguing nearby spheroidal object APPLES 1 \citep[][]{2005AJ....129..148P} may also fit in this category: its distance is uncertain, but its young age and likely isolation are consistent with recent quenching.

In the near future, various wide field surveys/instruments such as the Rubin Observatory's Legacy Survey of Space and Time \citep{2019ApJ...873..111I} and, later, the {\it Roman Space Telescope} (\citealt{2015arXiv150303757S}) should help us determine how common these quiescent isolated dwarfs are. The fact that COSMOS-dw1 was found in a very small and very well-studied field suggests that they may be quite common (as indicated by \citealt{2015MNRAS.454.1798K}) and can easily be missed.
Low surface brightness-optimized surveys, such as the HSC-SSP \citep{2018PASJ...70S...4A} and the Dragonfly Wide Field Survey \citep{2020ApJ...894..119D}, will provide an additional avenue
to obtaining a census of isolated low mass quiescent galaxies.

\acknowledgments
The authors would like to thank the anonymous referee for comments and suggestions that significantly improved the manuscript, as well as Dong Dong Shi and Xian Zhong Zheng for the initial identification of this galaxy. 
S.D. is supported by NASA through Hubble Fellowship grant HST-HF2-51454.001- A awarded by the Space Telescope Science Institute, which is operated by the Association of Universities for Research in Astronomy, Incorporated, under NASA contract NAS5-26555.
AJR was supported as a Research Corporation for Science Advancement Cottrell Scholar.

\bibliography{references}{}

\bibliographystyle{aasjournal}

\end{document}